\documentclass[%
    reprint,
    aps,prapplied,
    superscriptaddress,
    showkeys,
    longbibliography,
    amsmath,amssymb,
    balancelastpage]{revtex4-2}
\usepackage{mathptmx}
\usepackage{graphicx}   
\usepackage[none]{hyphenat}
\usepackage{siunitx}
\usepackage{ltxfront}
\usepackage{bm}
\usepackage{float}

\usepackage[colorlinks=true,
    linkcolor=blue,     
    citecolor=blue,     
    filecolor=blue,     
    urlcolor=blue]{hyperref}
\allowdisplaybreaks

\makeatletter
\newcommand\figref[1]{\@ifnextchar\bgroup{\figref@double{#1}}{\figref@single{#1}}}
\newcommand\figref@single[1]{\hyperref[{#1}]{\textup{Fig.~\ref*{#1}}}~}
\newcommand\figref@double[2]{\hyperref[{#1}]{\textup{Fig.~\ref*{#1}}\textcolor{blue}{(#2)}}}
\makeatother
\renewcommand{\eqref}[1]{\hyperref[{#1}]{\textup{(\ref*{#1})}}}

\begin{document}
\title{Nonreciprocal coupling enables tunable exceptional-point response in non-Hermitian circuits}

\author{Xiaoyin Sha}
\orcid{0009-0006-0180-2365}
\affiliation{\mbox{School of Electrical Engineering, Xi'an Jiaotong University, Xi'an 710049, China}}

\author{Chao Ma}
\orcid{0000-0003-3952-0516}
\email[Contact authors: ]{machao193032@xjtu.edu.cn}
\affiliation{\mbox{School of Microelectronics, Xi'an Jiaotong University, Xi'an 710049, China}}

\author{Yuqi Jiang}
\orcid{0009-0000-4523-6832}
\affiliation{\mbox{School of Electrical Engineering, Xi'an Jiaotong University, Xi'an 710049, China}}

\author{Zhiqin Jiang}
\orcid{0009-0005-7159-1156}
\affiliation{\mbox{School of Electrical Engineering, Xi'an Jiaotong University, Xi'an 710049, China}}

\author{Junan Lai}
\orcid{0009-0001-4990-7354}
\affiliation{\mbox{School of Electrical Engineering, Xi'an Jiaotong University, Xi'an 710049, China}}

\author{Xiaoke Gao}
\orcid{0000-0002-3769-4451}
\affiliation{\mbox{School of Electrical Engineering, Chongqing University, Chongqing 400044, China}}%

\author{Tianyu Dong}
\orcid{0000-0003-4816-0073}
\email[Contact authors: ]{tydong@mail.xjtu.edu.cn}
\affiliation{\mbox{School of Electrical Engineering, Xi'an Jiaotong University, Xi'an 710049, China}}

\date{October 29, 2026}

\begin{abstract}
    Exceptional points (EPs) exhibit characteristic square-root spectral responses to weak perturbations, with response magnitudes governed by configuration-dependent prefactors that are often relatively small. Modulating these prefactor while preserving the exceptional characteristics offers an additional degrees of freedom for engineering non-Hermitian sensing responses. Here, we report a nonreciprocal non-Hermitian circuit system with gyrator-meditated coupling that supports a continuous family of EP configurations. Unlike reciprocal coupling, which typically yields a unique EP configuration for balanced gain and loss, nonreciprocal coupling allows the directional forward and backward coupling strengths to be varied while maintaining their product and thus preserving the EP condition and yielding a library of EPs. This additional freedom enables continuous control of the frequency-splitting magnitude without modifying the square-root characteristic dependence on absolute perturbation strength. We show that the spectral splitting induced by a given absolute resistance perturbation can be enhanced by appropriately adjusting the coupling asymmetry. Theoretical predictions, circuit simulations, and experimental reflection-spectrum measurements consistently confirms this tunable response. Our results establish nonreciprocal coupling as an alternative approach for engineering and enhancing EP-induced perturbative responses without changing their characteristic branch-point scaling.
\end{abstract}

\keywords{Coupled oscillators, non-Hermitian circuit, non-reciprocal coupling, sensitivity enhancement, gyrator}

\maketitle 

\section{Introduction} \label{sec:intro}
Resonant sensing relies on converting weak perturbations into measurable spectral responses. Exceptional points (EPs) --- branch-point singularities of non-Hermitian systems at which both the eigenvalues and their associated eigenvectors coalesce --- offer a distinctive mechanism for producing nonanalytic spectral responses \cite{heiss2012physics,elganainy2018non}. Near an $N$th-order EP, a generic relative or absolute perturbation of strength $\varepsilon$ induces an eigenvalue splitting proportional to $\varepsilon^{1/N}$, i.e., $\Delta \omega \sim A\varepsilon^{1/N}$, where the characteristic fraction exponent $1/N$ is fixed by the order of the singularity and the prefactor $A$ is set by the underlying modal structure and couplings. This fractional-power scaling stands in contrast to the typically linear response of nondegenerate systems \cite{wiersig2014enhancing,wiersig2016sensors} and has motivated extensive investigations of EP-enhanced sensing in optical resonators \cite{hodaei2017enhanced,chen2017exceptional}, mechanical platforms \cite{kononchuk2022exceptional}, and electronics ranging from active gain--loss LRC circuits \cite{schindler2011active} to higher-order EP sensing in electronic and radio-frequency resonators \cite{xiao2019enhanced,yin2023highorder}. In addition, critical-point-based wireless LC sensors have also been demonstrated for detecting weak capacitive perturbations through reflection-spectrum measurements, highlighting the potential of non-Hermitian resonator systems for practical sensing applications \cite{ma2026critical}.

While the scaling exponent is protected by the EP order, the prefactor $A$ remains tunable. Quantitative measures of EP response strength and theoretical constraints on its magnitude have been established \cite{wiersig2022response,wiersig2022distance}, together with numerical approaches for extracting spectral response strength from wave simulations \cite{kullig2025spectral}. Independently controlling this prefactor while maintaining the EP condition thus provides an additional opportunity for engineering and enhancing the spectral response of non-Hermitian sensors \cite{wiersig2023moving,yang2025crosscoupling}. In reciprocal systems, the forward and backward coupling strengths are identical. For fixed gain and loss, the EP condition then fixes a unique coupling value, leaving little freedom to reshape the perturbative response without lifting the degeneracy. Nonreciprocal coupling relaxes this restriction \cite{ranzani2015graph,metelmann2015nonreciprocal,lau2018fundamental,xie2024quantum,zhao2024exceptional}: the EP condition can constrain a combination of the directional coupling strengths so that a continuous family of asymmetric configurations can support the same exceptional degeneracy. This additional freedom allows distinct asymmetric coupling configurations to support the same exceptional degeneracy, potentially enabling control of the perturbative response without shifting the system away from the EP.

Gyrators furnish a practical circuit-based approach to implementing a nonreciprocal coupling. Gyrator-based resonator systems have been employed to realize nonreciprocal exceptional degeneracies and investigate their singular perturbation responses \cite{rouhi2022exceptional,nikzamir2025exceptional}. In a balanced gain--loss system with asymmetric couplings, the EP condition can be maintained by fixing the product of the forward and backward coupling strengths while varying their ratio. The reciprocal configuration corresponds to a single point on this continuous manifold, whereas nonreciprocity enables access to a family of distinct EP configurations. This raises a fundamental question: Can coupling asymmetry be used to continuously tune the prefactor of the perturbative response while preserving the characteristic square-root scaling of a second-order EP?

Here, we exploit this freedom in a gyrator-coupled nonreciprocal gain--loss resonator. By varying the coupling asymmetry while preserving the product of the directional coupling strength, we realize a continuous family of second-order EP configurations. All members of the family exhibit the same characteristic square-root dependence on the perturbation strength, yet the corresponding frequency-splitting prefactor varies continuously with the coupling asymmetry. Increasing the coupling asymmetry enlarges the frequency splitting induced by a given small absolute resistance change, thereby improving the resolvability of minute resistance change under a fixed spectral resolution threshold. Our results identify nonreciprocal coupling as an additional control parameter that enhances the practical sensitivity of an EP sensor without altering its characteristic branch-point scaling.

\section{Gyrator-mediated nonreciprocal exceptional-point family}
\begin{figure}[!ht]
    \includegraphics[width=3.4in]{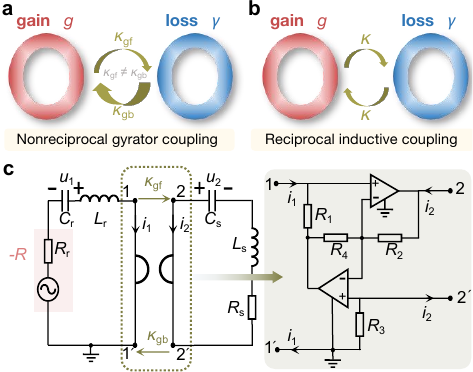}
    \caption{\textbf{Gyrator-mediated nonreciprocal coupling in a gain--loss resonator system}. (a) Nonreciprocal coupling mediated by a gyrator, with independently adjustable forward and backward coupling coefficients $\kappa_\text{gf}$ and $\kappa_\text{gb}$. (b) Conventional reciprocal inductive coupling between a gain resonator with gain rate $g$ and a loss resonator with loss rate $\gamma$, characterized by identical coupling strengths $\kappa$ in both directions. (c) Circuit implementation of the gyrator-coupled gain--loss resonator system. The gain and loss resonators are interconnected through an active gyrator, whose operational-amplifier implementation is shown in the right panel and consists of resistors $R_1$ -- $R_4$.}
    \label{fig:figure01}
\end{figure}

We consider a gain--loss resonator system with nonreciprocal systems mediated by a gyrator, as illustrated in \figref{fig:figure01}{a}. Unlike the conventional configuration, in which the resonators are inductively coupled and therefore exhibit identical forward and backward coupling strengths [\figref{fig:figure01}{b}], replacing the reciprocal inductive coupling with a gyrator enables independent control of the directional forward and backward coupling coefficients, $\kappa_\text{gf}$ and $\kappa_\text{gb}$, respectively. The corresponding circuit implementation, shown in \figref{fig:figure01}{c}, employs an operational-amplifier-based active gyrator characterized by the effective directional resistances $R_\text{gf} = -R_3$ and $R_\text{gb} = R_1 R_2/R_4$, which are referred to as transresistances. These independently adjustable resistances provide a means of controlling the coupling asymmetry, introducing an additional degree of freedom absent in the reciprocal system.

Applying Kirchhoff's voltage law to the coupled resonators yields
\begin{subequations} \label{eq:system-equation-2order-kvl}
    \begin{align}
        u_1 & = L_\text{r} \frac{\text{d}i_1}{\text{d}t} + R_\text{gf} i_2 - R i_1, \label{eq:system-equation-2order-kvl-u1}         \\
        u_2 & = L_\text{s}\frac{\text{d}i_2}{\text{d}t} + R_\text{gb} i_1 + R_\text{s} i_2, \label{eq:system-equation-2order-kvl-u2}
    \end{align}
\end{subequations}
where $u_n$ and $i_n$ ($n = 1, 2$) denote the capacitor voltages and resonator currents, respectively, with $i_1 = - C_\text{r} \text{d} u_1/\text{d} t$ and $i_2 = - C_\text{s} \text{d} u_2/\text{d} t$. For simplicity, we assume identical capacitances and inductances for the reader (subscript ``r'') and sensor (subscript ``s'') resonators, such that $C_\text{r} = C_\text{s} = C$ and $L_\text{r} = L_\text{s} = L$. The reader incorporates an effective negative resistance $-R$ to provide gain via a vector network analyzer (VNA), whereas the sensor contains a positive resistance $R_\text{s}$ representing dissipation. Assuming time-harmonic solutions with the convention $\text{exp}(\text{i}\omega t)$, the circuit equation \eqref{eq:system-equation-2order-kvl} can be expressed in matrix form as
\begin{equation} \label{eq:2order-CMT-parameters}
    \begin{pmatrix}
        - g + \text{i}(\omega / \omega_0 - \omega_0 / \omega) & - \kappa_\text{gf}                                       \\
        \kappa_\text{gb}                                      & \gamma + \text{i}(\omega / \omega_0 - \omega_0 / \omega)
    \end{pmatrix}
    \begin{pmatrix}
        i_1 \\ i_2
    \end{pmatrix}
    = 0,
\end{equation}
where $\omega_0 = 1 / \sqrt{LC}$ denotes the natural frequency of the LC oscillator. The dimensionless gain and loss parameters are defined as $g = R\sqrt{C/L}$ and $\gamma = R_\text{s}\sqrt{C/L}$, respectively, while $\kappa_\text{gf} = - R_\text{gf}\sqrt{C/L} = R_3\sqrt{C/L}$ and $\kappa_\text{gb} = R_\text{gb}\sqrt{C/L} = R_1 R_2/R_4\sqrt{C/L}$ represent the dimensionless forward and backward coupling coefficients induced by the gyrator.

\begin{figure}[!ht]
    \includegraphics[width=2.4in]{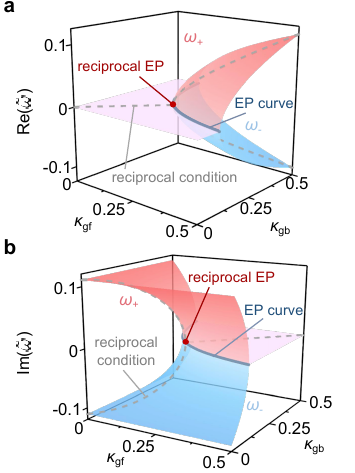}
    \caption{\textbf{Continuous family of exceptional-point configurations enabled by nonreciprocal coupling}. (a) Real and (b) imaginary parts of the normalized complex eigenfrequencies $\tilde{\omega}$ as functions of the directional coupling coefficients $\kappa_\text{gf}$ and $\kappa_\text{gb}$. The red and blue surfaces represent the two eigenfrequency branches $\omega_{+}$ and $\omega_{-}$, respectively. The dark-blue curve indicates the exceptional-point (EP) condition $\kappa_\text{gf} \kappa_\text{gb} = g^2$,  along which the eigenfrequencies coalesce. The gray dashed line denotes the reciprocal coupling, $\kappa_\text{gf} = \kappa_\text{gb}$, and its intersection with the EP curve (red point) identifies the reciprocal EP. Nonreciprocal coupling enables a continuous family of EP configurations along the same curve.}
    \label{fig:figure02}
\end{figure}

Under the near-resonance approximation $\omega / \omega_0 - 1 \approx 0$, the frequency-dependent term in \eqref{eq:2order-CMT-parameters} can be linearized as $\omega / \omega_0 - \omega_0 / \omega = (\omega^2 - \omega_0^2) / (\omega \omega_0) = (\omega - \omega_0) (\omega + \omega_0) / (\omega \omega_0) \approx 2 (\omega - \omega_0) / \omega_0$. Adopting the time harmonic convention $\exp(\text{i}\omega t)$, the resulting coupled-mode equations can be written in a Schr\"{o}dinger-like form as $\text{i}\text{d} \bm{i}/\text{d} t = \omega_0 \hat{H}_2 \bm{i}$ where $\bm{i} = (i_1, i_2)^\text{T}$ and
\begin{equation} \label{eq:2order-H}
    \hat{H}_\text{2}
    =
    \begin{pmatrix}
        - 1 + \text{i} g/2           & \text{i} \kappa_\text{gf}/2 \\
        - \text{i}\kappa_\text{gb}/2 & - 1 - \text{i} \gamma/2
    \end{pmatrix}
\end{equation}
is the normalized dimensionless effective Hamiltonian. Introducing the normalized frequency detuning $\tilde \omega = \omega / \omega_0 - 1$, the characteristic equation follows from $\det[\hat{H}_2 + (\tilde \omega + 1) \mathbb{I}_2] = 0$, yielding
\begin{equation} \label{eq:2order-characteristic-equation}
    \begin{split}
        \tilde \omega^2 + \frac{g \gamma}{4} - \frac{\kappa_\text{gf} \kappa_\text{gb}}{4} +  \frac{\text{i} \tilde \omega}{2} \left(g - \gamma \right) = 0,
    \end{split}
\end{equation}
where $\mathbb{I}_2$ denotes a second-order identity matrix. The corresponding complex frequency detunings are
\begin{equation} \label{eq:complex-eigen-frequency}
    \tilde{\omega}_{\pm} = -\frac{1}{4} \left[\text{i}(g - \gamma) \pm \sqrt{- \gamma^2 - g^2 - 2 \gamma g + 4 \kappa_\text{gb} \kappa_\text{gf}}\right].
\end{equation}
Second-order exceptional points occur when the two eigenfrequencies and their associated eigenvectors coalesce. For nonzero directional couplings, this condition is satisfied when the discriminant vanishes, giving a general EP condition $\kappa_\text{gf} \kappa_\text{gb} = (g+\gamma)^2/4$ and the complex eigenfrequency $\tilde{\omega}_\text{EP} = -\text{i}(g - \gamma)/4$ for an unbalanced system. For balanced gain and loss ($g=\gamma$), the EP condition reduces to
\begin{equation} \label{eq:EP-condition}
    \kappa_\text{gf} \kappa_\text{gb} = g^2,
\end{equation}
when the balanced system has a coalesced eigenfrequency at $\omega_0$, viz. $|\tilde \omega_{\pm}| = 0$. A key feature of \eqref{eq:EP-condition} is that it constrains only the product of the directional coupling coefficients, leaving their ratio adjustable. This property is illustrated by the complex-eigenfrequency surfaces in \figref{fig:figure02}, which show the real and imaginary parts of the two eigenfrequency branches as functions of $\kappa_\text{gf}$ and $\kappa_\text{gb}$. For fixed balanced gain and loss, the eigenfrequency branches coalesce along the continuous EP curve $\kappa_\text{gf}\kappa_\text{gb} = g^2$. Reciprocity imposes the additional constraint $\kappa_\text{gf} = \kappa_\text{gb}$, which intersects the EP curve at the unique positive-coupling configuration $\kappa_\text{gf} = \kappa_\text{gb} = \kappa = g$. Breaking reciprocity removes this constraint, allowing a continuous family of asymmetric coupling configurations to satisfy the same EP condition.

To characterize the coupling asymmetry along this family, we introduce the dimensionless nonreciprocity parameter
\begin{equation}
    \eta = {\kappa_\text{gf}}/{\kappa_\text{gb}}.
\end{equation}
The reciprocal configuration corresponds to $\eta = 1$, whereas $\eta\neq1$ describe nonreciprocal coupling. Combining this definition with the balanced gain--loss EP condition \eqref{eq:EP-condition} gives $\kappa_\text{gf} = g\sqrt{\eta}$ and $\kappa_\text{gb} = g/\sqrt{\eta}$ for positive directional coupling coefficients. Thus, the coupling product $\kappa_\text{gf}\kappa_\text{gb}$ determines the EP condition, while the coupling ratio specifies the asymmetry between the two directions. Varying $\eta$ continuously reconfigures the directional coupling strengths without lifting the exceptional degeneracy. As demonstrated below, this additional freedom enables the magnitude of the perturbation-induced frequency splitting to be tuned while preserving the characteristic square-root scaling of a second-order EP.

\section{Nonreciprocity-controlled exceptional-point response}
The continuous family of nonreciprocal EP configurations provides an additional degree of freedom for controlling the perturbative response. Although all configurations along this family satisfy the same EP condition, their directional coupling strengths differ according to the nonreciprocity parameter $\eta$. Consequently, identical perturbations can produce different eigenfrequency splitting even when the unperturbed eigenvalues remain degenerated.

\begin{figure}[!ht]
    \includegraphics[width=3in]{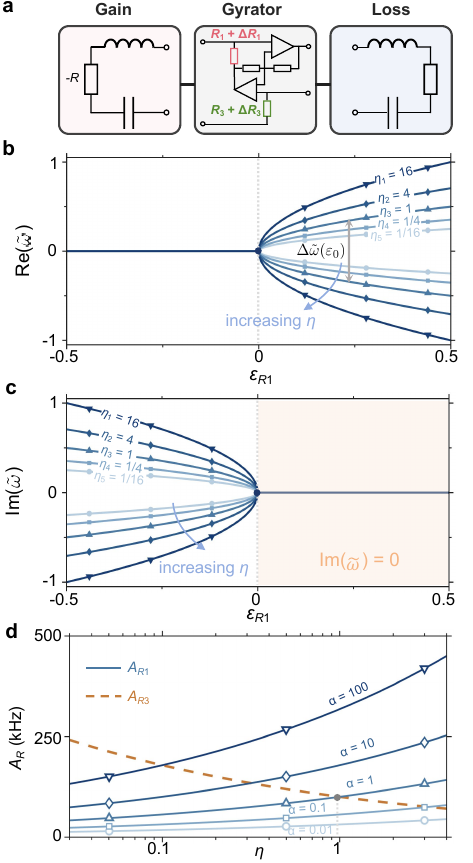}
    \caption{\textbf{Nonreciprocity-controlled eigenfrequency response and tunability of the exceptional-point response coefficient.} (a) Gain--gyrator--loss resonator system with separately applied resistance perturbations $\Delta R_1$ or $\Delta R_3$, which modifies the backward and forward coupling coefficients, respectively. (b) Real parts of the normalized eigenfrequencies $\tilde{\omega}_\pm$ as functions of the normalized absolute resistance perturbation $\varepsilon_{R1}$ for different coupling asymmetries $\eta = \kappa_\text{gf}/\kappa_\text{gb}$ of $\eta_1 = 16$, $\eta_2 = 4$, $\eta_3 = 1$, $\eta_4 = 1/4$, $\eta_5 = 1/16$. All configurations exhibit the characteristic square-root branch splitting of a second-order exceptional point, while the magnitude of the splitting varies systematically with $\eta$. (c) Imaginary parts of the normalized eigenfrequencies for the same configurations. For positive perturbations and positive directional coupling coefficients, the imaginary parts remain zero, indicating that the split eigenfrequencies stay on the real-frequency axis in this perturbation regime. (d) Theoretical square-root response coefficients $A_{R1}$ and $A_{R3}$ as functions of the coupling asymmetry $\eta$. The blue solid curves correspond to $R_1$ perturbations for gyrator resistance ratios $\alpha = R_2/R_4 = 0.01$, 0.1, 1, 10, and 100, whereas the orange dashed curve corresponds to $R_3$ perturbations, for which the response is independent of $\alpha$ at fixed gain and loss.}
    \label{fig:figure03}
\end{figure}

\paragraph{Resistance perturbation applied to $R_1$.}
To investigate this effect, we introduce a resistance perturbation $\Delta R_1$ into the gyrator, as indicated in red in \figref{fig:figure03}{a}. Since $R_\text{gb} = \alpha R_1$, where $\alpha = R_2/R_4$ is held constant, varying $R_1$ modifies the backward coupling $\kappa_\text{gb}$ while leaving $\kappa_\text{gf}$ unchanged. By defining the normalized resistance perturbation as $\varepsilon_{R1} = \Delta R_1 \sqrt{C/L}$. The perturbed effective Hamiltonian is therefore
\begin{equation} \label{eq:method-perturbed-H1}
    \hat{H}_2^{R1} = \begin{pmatrix}
        -1+\text{i}g/2                                          & \text{i} \kappa_\text{gf}/2 \\[4pt]
        -\text{i}(\kappa_\text{gb} + \alpha \varepsilon_{R1})/2 & - 1 - \text{i}\gamma/2
    \end{pmatrix}.
\end{equation}
Under the balanced gain--loss condition $g = \gamma$, the characteristic equation $\det[\hat{H}_2^{R1} + (\tilde \omega + 1) \mathbb{I}_2] = 0$ becomes
\begin{equation}
    \tilde{\omega}^{2} + \frac{g^2 - \kappa_\text{gf}(\kappa_\text{gb} + \alpha \varepsilon_{R1})}{4} = 0.
\end{equation}
Using the unperturbed EP condition $\kappa_\text{gf} \kappa_\text{gb} = g^2$, we obtain $\tilde{\omega}^{2} - \kappa_\text{gf} \alpha \varepsilon_{R1}/4 = 0$. The perturbed eigenfrequencies are thus $\tilde{\omega}_{\pm} = \pm \sqrt{\kappa_\text{gf} \alpha \varepsilon_{R1}}/2$, yielding the normalized eigenfrequency splitting
\begin{equation} \label{eq:method-normalized-splitting}
    \Delta\tilde{\omega}_{R1} = \left| \tilde{\omega}_{+} - \tilde{\omega}_{-} \right| = \sqrt{\kappa_\text{gf}\alpha \varepsilon_{R1}}.
\end{equation}
It is evident that the characteristic square-root dependence on perturbation strength is preserved, while the splitting amplitude is governed by the unperturbed coupling coefficient in the opposite direction.

\figref{fig:figure03}{b} and \figref{fig:figure03}{c} show the real and imaginary parts of the normalized eigenfrequency for different coupling asymmetries. All configurations exhibit the characteristic square-root branch splitting of a second-order EP, but the separation between the real-eigenfrequency branches varies systematically with $\eta$. For positive perturbations and positive directional couplings, the eigenfrequency detunings remain real. Thus, varying the coupling asymmetry changes the magnitude of the spectral response without altering its square-root dependence on perturbation strength. Since the normalized eigenfrequency reads $\tilde{\omega} = \omega/\omega_0 - 1$, the eigenfrequency splitting $\Delta\tilde{\omega} = |\tilde{\omega}_+ - \tilde{\omega}_-|$ is related to the physical frequency splitting by $\Delta f = |\omega_+ - \omega_-|/2\pi = f_0 \Delta \tilde{\omega}$, where $f_0 = \omega_0/(2\pi)$ is the uncoupled natural resonance frequency. The dependence of the response strength on nonreciprocity can therefore be obtained explicitly from the EP parameterization. Substituting $\kappa_\text{gf} = g \sqrt{\eta}$ into \eqref{eq:method-normalized-splitting}, the frequency splitting in physical units becomes
\begin{equation} \label{eq:sqrt-response}
    \Delta f_{R1} = A_{R1}(\eta) \sqrt{\varepsilon_{R1}},
\end{equation}
where
\begin{equation} \label{eq:A-eta}
    A_{R1}(\eta) = f_0 \sqrt{g \alpha} \eta^{1/4},
\end{equation}
Here, \eqref{eq:A-eta} demonstrates that the square-root exponent is fixed by the second-order EP degeneracy, whereas the response coefficient (prefactor) follows a quarter-power dependence on the coupling asymmetry, provided that $g$ and $\alpha$ remain fixed, as illustrated by the blue solid curves in \figref{fig:figure03}{d}. Moreover, for a given reciprocity $\eta$, the response prefactor scales as $A_{R1}\propto\sqrt{\alpha}$, providing an additional degree of freedom for controlling the magnitude of the spectral response. Together, these dependencies demonstrate that the frequency splitting induced by a given resistance perturbation can be continuously tailored through both coupling asymmetry and the gyrator resistance ratio, without altering the EP degeneracy or its characteristic square-root scaling.

\paragraph{Resistance perturbation applied to $R_3$.}
We also consider a resistance perturbation $\Delta R_3$ applied to the gyrator, as indicated in green in \figref{fig:figure03}{a}. Since $R_\text{gf} = -R_3$, this perturbation modifies the forward coupling $\kappa_\text{gf}$ while leaving the backward coupling strength $\kappa_\text{gb}$ unchanged. Defining the normalized resistance perturbation as $\varepsilon_{R3} = \Delta R_3 \sqrt{C/L}$, the perturbed effective Hamiltonian becomes
\begin{equation} \label{eq:method-perturbed-H3}
    \hat{H}_2^{R3} = \begin{pmatrix}
        -1+\text{i}g/2              & \text{i} (\kappa_\text{gf} + \varepsilon_{R3})/2 \\[4pt]
        -\text{i}\kappa_\text{gb}/2 & - 1 - \text{i}\gamma/2
    \end{pmatrix},
\end{equation}
yielding the normalized frequency splitting $\Delta\tilde{\omega}_{R3} = \sqrt{\kappa_\text{gb} \varepsilon_{R3}}$ and in physical unit
\begin{equation} \label{eq:sqrt-response3}
    \Delta f_{R3} = A_{R3}(\eta) \sqrt{\varepsilon_{R3}},
\end{equation}
where
\begin{equation} \label{eq:A-eta-R3}
    A_{R3}(\eta) = f_0\sqrt{g} \eta^{-1/4}.
\end{equation}
Equation \eqref{eq:A-eta-R3} shows that the response factor $A_{R3}$ follows an inverse quarter-power dependence on the coupling asymmetry $\eta$ for fixed gain $g$, as illustrated by the orange dashed curve in \figref{fig:figure03}{d}. This behavior contrasts with that of the $R_1$ perturbation, for which the response prefactor increases with $\eta$. Both perturbations preserve the characteristic square-root scaling of a second-order EP, while exhibiting opposite dependencies of the response coefficient on the coupling asymmetry. These results demonstrate that nonreciprocity enables direction-dependent control of the spectral response, allowing the same EP configuration to exhibit different frequency-splitting amplitudes depending on which directional coupling is perturbed.

\begin{figure}[!ht]
    \includegraphics[width=3.2in]{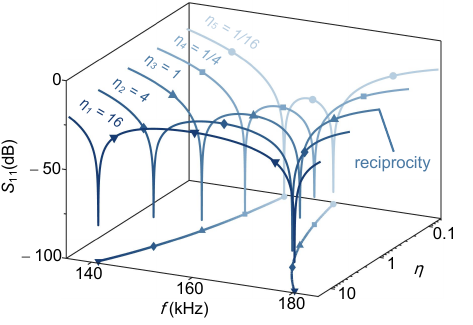}
    \caption{\textbf{Reflection-spectrum signatures of the nonreciprocity-controlled EP response}. Reflection coefficient $S_{11}$ as a function of frequency $f$ and the coupling-asymmetry parameter $\eta$ under a fixed resistance perturbation. The trace at $\eta = 1$ corresponds to the reciprocal configuration, whereas the other traces correspond to nonreciprocal EP configurations along the unperturbed EP manifold. Here, $\alpha = 1$ is considered.}
    \label{fig:figure04}
\end{figure}

\paragraph{Reflection-spectrum characterization and experimental validation.}
We validate the nonreciprocity-dependent eigenfrequency response by perturb $R_1$. As shown in \figref{fig:figure04}, the reflection coefficient $S_{11}$ exhibits distinct resonance features whose separation varies continuously with the coupling asymmetry $\eta$ under a fixed resistance perturbation. The reciprocal configuration corresponds to $\eta = 1$, while the other traces represent different nonreciprocal EP configurations before perturbation. As $\eta$ is varied along the unperturbed EP manifold, the spectral separation changes continuously, consistent with the predicted tunability of the eigenfrequency splitting, that is, larger $\eta$ produces a larger resonance splitting (consistent with the corresponding increase in the absolute response strength). These results demonstrate that the nonreciprocity-controlled perturbative response can be resolved in the reflection spectrum, providing a spectroscopic signature of the tunable EP response.

\begin{figure*}
    \includegraphics[width=0.92\textwidth]{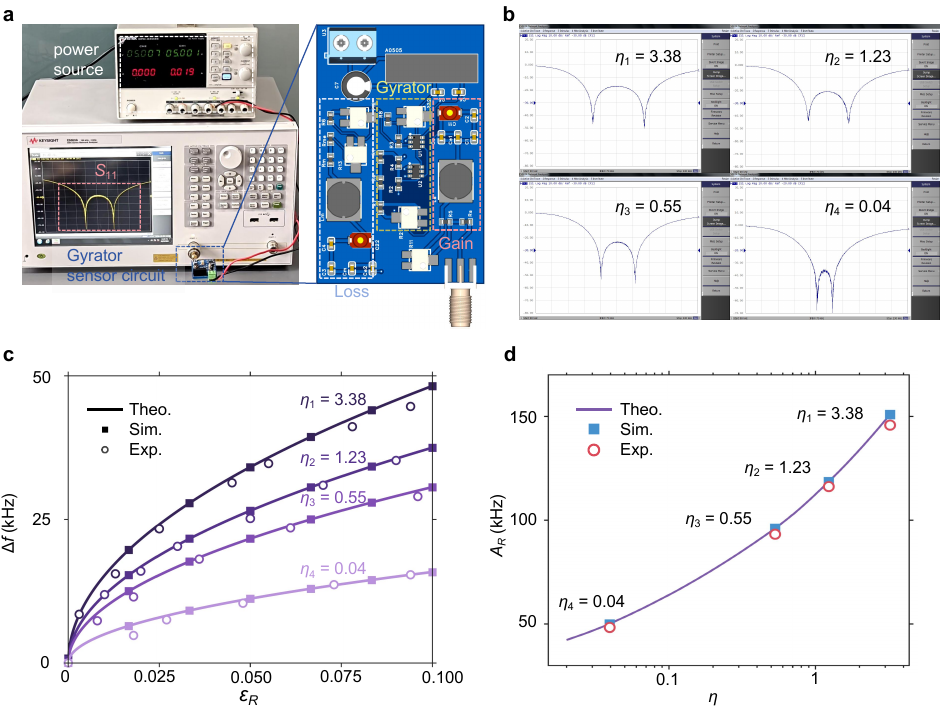}
    \caption{\textbf{Experimental validation of the nonreciprocity-controlled exceptional-point response.} (a) Experimental setup for measuring the reflection response of the gyrator--coupled gain--loss resonator system, showing the vector network analyzer (VNA), power source, and gyrator sensor circuit, together with an enlarged view of the circuit implementation. (b) Representative VNA screenshots showing the measured reflection coefficient $S_{11}$ for four nonreciprocal EP configurations with $\eta = 3.38$, 1.23, 0.55, and 0.04. (c) Frequency splitting $\Delta f$ as a function of the normalized resistance perturbation $\varepsilon_{R1}$ for the four EP configurations. Solid curves, filled squares, and open circles represent theoretical predictions, LTspice simulations, and experimental measurements, respectively. (d) Square-root response coefficient $A$, defined by $\Delta f = A_R(\eta)\sqrt{\varepsilon_R}$, as a function of the coupling-asymmetry parameter $\eta$. The solid curve, filled squares, and open circles denote the theoretical prediction, LTspice simulations, and experimental measurements, respectively.}
    \label{fig:figure05}
\end{figure*}

The predicted nonreciprocity-controlled response is validated through circuit simulations and experiments, as summarized in \figref{fig:figure05}. The experimental setup is shown in \figref{fig:figure05}{a}, and representative measured reflection spectra for different coupling asymmetries are presented in \figref{fig:figure05}{b}. Both resonators have identical inductances $L_1 = L_2 = 100~\si{\micro\henry}$ and capacitances $C_1 = C_2 = 10~\si{nF}$, corresponding to an uncoupled natural resonance frequency $f_0 = 159~ \si{kHz}$. In both simulations and the experiments, the gyrator is implemented using two ADA4807 operational amplifiers powered by $\pm 5~\si{V}$, with $R_2 = R_4 = 100~\si{\ohm}$. Accounting for the finite internal resistance of the practical circuit, the EP condition is approximately satisfied at balanced gain and loss parameters $g = \gamma = 0.4$. Four representative nonreciprocal EP configurations, characterized by $\eta = 3.38, 1.23, 0.55, 0.04$, are realized with the resistor pairs $(R_1, R_3) =(21, 71)~\si{\ohm}, (35, 43)~\si{\ohm}, (51, 28)~\si{\ohm}$, and $(203, 7.3)~\si{\ohm}$, respectively.

Under the same resistance perturbation, distinct EP configurations exhibit different separations between their resonance features, demonstrating the tunability of the spectral response through nonreciprocal coupling. \figref{fig:figure05}{c} quantitatively compares the frequency splitting $\Delta f$ as a function of the normalized resistance perturbation $\varepsilon_R$ for four representative values of $\eta$. The theoretical predictions, LTspice simulations, and experimental measurements show consistent behavior, with all configurations exhibiting the characteristic square-root dependence on perturbation strength but different splitting amplitudes. Despite the different coupling asymmetries, the response follows the same fractional-power scaling, while the magnitude of the frequency splitting varies substantially among the EP configurations. To quantify this variation, we characterize each EP configuration by the square-root response coefficient $A$ defined by $\Delta f = A(\eta) \sqrt{\varepsilon_R}$. As shown in \figref{fig:figure05}{d}, $A$ increases systematically with $\eta$, in agreement with the theoretical prediction $A(\eta)\propto\eta^{1/4}$ derived in \eqref{eq:A-eta}. The agreement among theory, simulation, and experiment demonstrates that nonreciprocal coupling enables continuous control of the frequency-splitting amplitude while preserving the characteristic square-root scaling of a second-order EP.

The observed response tunability originates from the different roles of the directional coupling product and ratio in nonreciprocal EP systems. The coupling product fixes the exceptional degeneracy, whereas the coupling ratio controls the distribution of interaction strengths between the two directions. A perturbation applied to one coupling channel therefore produces a splitting that depends on the complementary directional coupling, allowing different EP configurations to exhibit distinct response coefficients while retaining identical square-root scaling. We stress that perturbations considered correspond to the absolute resistance changes rather than relative variations. Consequently, an appropriate choice of the initial gyrator resistance configuration can enhance the frequency splitting induced by a given small absolute resistance perturbation. This separation establishes coupling asymmetry as an additional parameter for engineering perturbative spectral responses without altering the underlying exceptional degeneracy.

\section{Conclusions}
We have demonstrated that nonreciprocal coupling enables continuous control of the perturbative spectral response of exceptional-point (EP) systems without lifting the exceptional degeneracy. In a gyrator-coupled gain--loss resonator system, the EP condition constrains the product of the directional coupling coefficients, while their ratio remains adjustable. This separation gives rise to a continuous family of EP configurations with identical coalesced eigenfrequencies but different coupling asymmetries. The adjustable gyrator resistances provide a practical circuit-level approach to realizing and reconfiguring these EP states. When a resistance perturbation is applied to one coupling channel, the resulting eigenfrequency splitting depends on the complementary directional coupling, allowing the response amplitude to be tuned while preserving the characteristic square-root scaling. The predicted behavior is supported by analytical theory, LTspice simulations, and experimental reflection measurements. These findings establish coupling asymmetry as an additional design parameter for engineering perturbative responses near exceptional degeneracies. Although the enhanced spectral response does not necessarily translate into improved detection limits \cite{langbein2018exceptional,anderson2023clarification}, the ability to tune the splitting coefficient independently of the EP condition offers a flexible approach to controlling non-Hermitian resonator dynamics. This strategy provides opportunities for designing electronic systems with adjustable spectral responses for sensing and signal-processing applications. A promising extension of the proposed framework lies in wireless sensing and telemetry, where non-Hermitian sensor telemetry and electronically tunable nonreciprocal EP architectures have attracted growing interest \cite{chen2018generalized,hao2026wireless}. Integrating coupling-asymmetry control into such platforms could enable in situ tuning of the spectral response strength while maintaining EP condition, providing a route toward reconfigurable wireless sensing systems.

\section*{Appendix: Methods}
\paragraph{Reflection-spectrum calculation.}
The experimentally accessible spectral response is characterized by the one-port scattering parameter $S_{11}$. Under the time-harmonic convention $\exp(i\omega t)$, the impedances of the reader and sensor resonators are $Z_\text{r}(\omega) = -R + \text{i} \omega L + 1/\text{i}\omega C,$ and $Z_\text{s}(\omega) = R_\text{s} + \text{i} \omega L + 1/\text{i}\omega C,$ respectively, where $-R$ represents the effective negative resistance of the reader and $R_\text{s}$ accounts for dissipation in the sensor. To determine the input impedance seen from the reader port, we eliminate the sensor current using the coupled circuit equations. The sensor equation gives $i_2 = - i_1 R_\text{gb}/Z_\text{s}(\omega)$. Substituting this relation into the reader equation yields $Z_\text{in}(\omega) = Z_\text{r}(\omega) - R_\text{gf}R_\text{gb}/Z_\text{s}(\omega)$, where $R_\text{gf}$ and $R_\text{gb}$ are the signed directional transresistances of the gyrator. For a measurement port with characteristic impedance $Z_0$, the reflection coefficient reads
\begin{equation} \label{eq:method-S11}
    S_{11}(\omega) = \frac{Z_\text{in}(\omega)-Z_0}{Z_\text{in}(\omega) + Z_0},
\end{equation}
which is often expressed as $S_{11,\text{dB}} = 20 \log_{10} \left| S_{11} \right|$ in decibels.

\paragraph{Circuit simulation and experimental data analysis.}
Circuit simulations were performed in LTspice using the gain--gyrator--loss topology described by the analytical model. For each nonreciprocity parameter $\eta$, the directional couplings were adjusted to satisfy the unperturbed EP condition. A resistance perturbation $\Delta R_1$ was then applied to the gyrator, with the remaining circuit parameters held fixed for that configuration. The resulting reflection spectra were analyzed to extract the frequencies of the two split resonance features. Experimentally, the spectral response was characterized by measuring the one-port reflection coefficient $S_{11}$ from the reader side of the coupled resonator circuit. Different nonreciprocal EP configurations were established by adjusting the gyrator resistances while maintaining the EP condition. For each configuration, a controlled resistance perturbation $\Delta R_1$ was introduced, and the corresponding reflection spectrum was recorded. The frequencies $f_{+}$ and $f_{-}$ of the two resolved resonance features were extracted, and their separation was calculated as $\Delta f = \left| f_{+} - f_{-} \right|$. To quantify the perturbative response, the frequency splitting for each coupling configuration $\eta$ was fitted to the square-root relation $\Delta f = A(\eta) \sqrt{\varepsilon_R}$, where $\varepsilon_R = \Delta R_1\sqrt{C/L}$ is the normalized resistance perturbation. Assuming that the splitting vanishes at the unperturbed EP, the least-squares fit was constrained to pass through the origin. Minimizing the sum of squared residuals gives $A = \sum\limits_n \sqrt{\varepsilon_{R,n}} \Delta f_n /\sum\limits_n\varepsilon_{R,n}$, where $\Delta f_n$ is the extracted frequency splitting at perturbation strength $\varepsilon_{R,n}$. The same fitting procedure was applied to the analytical predictions, LTspice simulations, and experimental measurements, enabling a consistent comparison of the response coefficient $A$ across different nonreciprocal EP configurations.

\paragraph*{Acknowledgments:} X. Sha thanks Zhuoyu Zhang for helpful discussion.

\paragraph*{Funding:} T. Dong acknowledges the support of the National Natural Science Foundation of China (52577015).

\paragraph*{Data and code availability:} All data needed to evaluate the conclusions are presented in the article. Raw data, corresponding simulation data, and the relevant code are available upon reasonable request.

\paragraph*{Competing interests:} The authors declare that they have no competing interests.

\bibliography{refs}

@article{anderson2023clarification,
    title = {Clarification of the Exceptional-Point Contribution to Photonic Sensing},
    author = {Anderson, Dalton and Shah, Manav and Fan, Linran},
    journal = {Physical Review Applied},
    volume = {19},
    number = {3},
    pages = {034059},
    year = {2023},
    doi = {10.1103/PhysRevApplied.19.034059}
}

@article{chen2017exceptional,
    title = {Exceptional points enhance sensing in an optical microcavity},
    author = {Chen, Weijian and {\"O}zdemir, {\c{S}}ahin Kaya and Zhao, Guangming and Wiersig, Jan and Yang, Lan},
    journal = {Nature},
    volume = {548},
    number = {7666},
    pages = {192--196},
    year = {2017},
    doi = {10.1038/nature23281}
}

@article{chen2018generalized,
    title = {Generalized parity--time symmetry condition for enhanced sensor telemetry},
    author = {Chen, Pai-Yen and Sakhdari, Maryam and Hajizadegan, Mehdi and Cui, Qingsong and Cheng, Mark Ming-Cheng and El-Ganainy, Ramy and Al{\`u}, Andrea},
    journal = {Nature Electronics},
    volume = {1},
    number = {5},
    pages = {297--304},
    year = {2018},
    doi = {10.1038/s41928-018-0072-6}
}

@article{elganainy2018non,
    title = {Non-{Hermitian} physics and {PT} symmetry},
    author = {El-Ganainy, Ramy and Makris, Konstantinos G. and Khajavikhan, Mercedeh and Musslimani, Ziad H. and Rotter, Stefan and Christodoulides, Demetrios N.},
    journal = {Nature Physics},
    volume = {14},
    number = {1},
    pages = {11--19},
    year = {2018},
    doi = {10.1038/nphys4323}
}

@article{hao2026wireless,
    title = {An exceptional-point wireless sensor featuring non-reciprocal and electronic-tunable coupling},
    author = {Hao, Yushun and Zhang, Xuanhao and Wu, Pengde and Wang, Gaofeng and Cheng, Yuhua},
    journal = {Applied Physics Letters},
    volume = {128},
    number = {19},
    pages = {193304},
    year = {2026},
    doi = {10.1063/5.0333910}
}

@article{heiss2012physics,
    title = {The physics of exceptional points},
    author = {Heiss, W. D.},
    journal = {Journal of Physics A: Mathematical and Theoretical},
    volume = {45},
    number = {44},
    pages = {444016},
    year = {2012},
    doi = {10.1088/1751-8113/45/44/444016}
}

@article{hodaei2017enhanced,
    title = {Enhanced sensitivity at higher-order exceptional points},
    author = {Hodaei, Hossein and Hassan, Absar U. and Wittek, Steffen and Garcia-Gracia, Hipolito and El-Ganainy, Ramy and Christodoulides, Demetrios N. and Khajavikhan, Mercedeh},
    journal = {Nature},
    volume = {548},
    number = {7666},
    pages = {187--191},
    year = {2017},
    doi = {10.1038/nature23280}
}

@article{kononchuk2022exceptional,
    title = {Exceptional-point-based accelerometers with enhanced signal-to-noise ratio},
    author = {Kononchuk, Rodion and Cai, Jizhe and Ellis, Fred and Thevamaran, Ramathasan and Kottos, Tsampikos},
    journal = {Nature},
    volume = {607},
    number = {7920},
    pages = {697--702},
    year = {2022},
    doi = {10.1038/s41586-022-04904-w}
}

@article{kullig2025spectral,
    title = {Calculating the spectral response strength of non-{Hermitian} systems with an exceptional point directly from wave simulations},
    author = {Kullig, Julius and Wiersig, Jan},
    journal = {Physical Review Research},
    volume = {7},
    number = {1},
    pages = {013223},
    year = {2025},
    doi = {10.1103/PhysRevResearch.7.013223}
}

@article{langbein2018exceptional,
    title = {No exceptional precision of exceptional-point sensors},
    author = {Langbein, Wolfgang},
    journal = {Physical Review A},
    volume = {98},
    number = {2},
    pages = {023805},
    year = {2018},
    doi = {10.1103/PhysRevA.98.023805}
}

@article{lau2018fundamental,
    title = {Fundamental limits and non-reciprocal approaches in non-{Hermitian} quantum sensing},
    author = {Lau, Hoi-Kwan and Clerk, Aashish A.},
    journal = {Nature Communications},
    volume = {9},
    number = {1},
    pages = {4320},
    year = {2018},
    doi = {10.1038/s41467-018-06477-7}
}

@article{ma2026critical,
    title = {Critical point-based wireless sensors enabling tiny perturbation detection},
    author = {Ma, Chao and Yin, Ke and Zhang, Zhuoyu and Huang, Ruyi and Huang, Yuangen and Gan, Lanyue and Liu, Yuxuan and Xia, Fan and Xia, Tian and Chen, Yufan and Li, Wangchang and Peng, Lian-Mao and Dong, Tianyu and Hu, Youfan},
    journal = {Science Advances},
    volume = {12},
    number = {21},
    pages = {eaea6541},
    year = {2026},
    doi = {10.1126/sciadv.aea6541}
}

@article{metelmann2015nonreciprocal,
    title = {Nonreciprocal Photon Transmission and Amplification via Reservoir Engineering},
    author = {Metelmann, A. and Clerk, A. A.},
    journal = {Physical Review X},
    volume = {5},
    number = {2},
    pages = {021025},
    year = {2015},
    doi = {10.1103/PhysRevX.5.021025}
}

@article{nikzamir2025exceptional,
    title = {Exceptional points in a gyrator-based circuit and high-sensitivity oscillator for nonlinear applications},
    author = {Nikzamir, Alireza and Rouhi, Kasra and Figotin, Alexander and Capolino, Filippo},
    journal = {Physical Review Applied},
    volume = {24},
    number = {4},
    pages = {044054},
    year = {2025},
    doi = {10.1103/6t64-gm44}
}

@article{ranzani2015graph,
    title = {Graph-based analysis of nonreciprocity in coupled-mode systems},
    author = {Ranzani, Leonardo and Aumentado, Jos{\'e}},
    journal = {New Journal of Physics},
    volume = {17},
    number = {2},
    pages = {023024},
    year = {2015},
    doi = {10.1088/1367-2630/17/2/023024}
}

@article{rouhi2022exceptional,
    title = {Exceptional point in a degenerate system made of a gyrator and two unstable resonators},
    author = {Rouhi, Kasra and Nikzamir, Alireza and Figotin, Alexander and Capolino, Filippo},
    journal = {Physical Review A},
    volume = {105},
    number = {3},
    pages = {032214},
    year = {2022},
    doi = {10.1103/PhysRevA.105.032214}
}

@article{schindler2011active,
    title = {Experimental study of active {LRC} circuits with {PT} symmetries},
    author = {Schindler, Joseph and Li, Ang and Zheng, Mei C. and Ellis, F. M. and Kottos, Tsampikos},
    journal = {Physical Review A},
    volume = {84},
    number = {4},
    pages = {040101(R)},
    year = {2011},
    doi = {10.1103/PhysRevA.84.040101}
}

@article{wiersig2014enhancing,
    title = {Enhancing the Sensitivity of Frequency and Energy Splitting Detection by Using Exceptional Points: Application to Microcavity Sensors for Single-Particle Detection},
    author = {Wiersig, Jan},
    journal = {Physical Review Letters},
    volume = {112},
    number = {20},
    pages = {203901},
    year = {2014},
    doi = {10.1103/PhysRevLett.112.203901}
}

@article{wiersig2016sensors,
    title = {Sensors operating at exceptional points: General theory},
    author = {Wiersig, Jan},
    journal = {Physical Review A},
    volume = {93},
    number = {3},
    pages = {033809},
    year = {2016},
    doi = {10.1103/PhysRevA.93.033809}
}

@article{wiersig2022distance,
    title = {Distance between exceptional points and diabolic points and its implication for the response strength of non-{Hermitian} systems},
    author = {Wiersig, Jan},
    journal = {Physical Review Research},
    volume = {4},
    number = {3},
    pages = {033179},
    year = {2022},
    doi = {10.1103/PhysRevResearch.4.033179}
}

@article{wiersig2022response,
    title = {Response strengths of open systems at exceptional points},
    author = {Wiersig, Jan},
    journal = {Physical Review Research},
    volume = {4},
    number = {2},
    pages = {023121},
    year = {2022},
    doi = {10.1103/PhysRevResearch.4.023121}
}

@article{wiersig2023moving,
    title = {Moving along an exceptional surface towards a higher-order exceptional point},
    author = {Wiersig, Jan},
    journal = {Physical Review A},
    volume = {108},
    number = {3},
    pages = {033501},
    year = {2023},
    doi = {10.1103/PhysRevA.108.033501}
}

@article{xiao2019enhanced,
    title = {Enhanced Sensing and Nondegraded Thermal Noise Performance Based on {PT}-Symmetric Electronic Circuits with a Sixth-Order Exceptional Point},
    author = {Xiao, Zhicheng and Li, Huanan and Kottos, Tsampikos and Al{\`u}, Andrea},
    journal = {Physical Review Letters},
    volume = {123},
    number = {21},
    pages = {213901},
    year = {2019},
    doi = {10.1103/PhysRevLett.123.213901}
}

@article{xie2024quantum,
    title = {Quantum sensing with nonreciprocal couplings},
    author = {Xie, Dong and Xu, Chunling},
    journal = {Physical Review Applied},
    volume = {22},
    number = {6},
    pages = {064072},
    year = {2024},
    doi = {10.1103/PhysRevApplied.22.064072}
}

@article{yang2025crosscoupling,
    title = {Cross-coupling-induced sensitivity enhancement beyond divergent exceptional points},
    author = {Yang, Minye and Jian, Baolong and Ye, Zhilu and Wang, Lukang and Liu, Ming},
    journal = {Physical Review Applied},
    volume = {23},
    number = {6},
    pages = {064026},
    year = {2025},
    doi = {10.1103/PhysRevApplied.23.064026}
}

@article{yin2023highorder,
    title = {High-Order Exceptional Points in {Pseudo-Hermitian} Radio-Frequency Circuits},
    author = {Yin, Ke and Hao, Xianglin and Huang, Yuangen and Zou, Jianlong and Ma, Xikui and Dong, Tianyu},
    journal = {Physical Review Applied},
    volume = {20},
    number = {2},
    pages = {L021003},
    year = {2023},
    doi = {10.1103/PhysRevApplied.20.L021003}
}

@article{zhao2024exceptional,
    title = {Exceptional points induced by unidirectional coupling in electronic circuits},
    author = {Zhao, Wenzheng and Zhang, Yang and Gao, Zixuan and Peng, Delong and Kou, Jun-Long and Lu, Yan-Qing and El-Ganainy, Ramy and {\"O}zdemir, {\c{S}}ahin K. and Zhong, Qi},
    journal = {Nature Communications},
    volume = {15},
    number = {1},
    pages = {9907},
    year = {2024},
    doi = {10.1038/s41467-024-53929-4}
}

\end{document}